\documentclass[10pt,twocolumn,letterpaper]{article}

\usepackage{iccv}              
\usepackage{multirow}
\usepackage[accsupp]{axessibility}

\definecolor{iccvblue}{rgb}{0.21,0.49,0.74}
\usepackage[pagebackref,breaklinks,colorlinks,allcolors=iccvblue]{hyperref}

\def\paperID{*****} 
\def\confName{ICCV}
\def\confYear{2025}

\title{NeurOp-Diff: Continuous Remote Sensing Image Super-Resolution via Neural Operator Diffusion}

\author{Zihao Xu$^{1,2}$, Yuzhi Tang$^{1,2}$\thanks{Corresponding Author.} , Bowen Xu$^{1,2}$, Qingquan Li$^{1,2}$ \\
$^1$Guangdong Laboratory of Artificial Intelligence and Digital Economy (SZ), China \\
$^2$ Shenzhen University, China \\
{\tt\small \{xuzihao, tangyuzhi\}@gml.ac.cn, \{bowenxu, liqq\}@szu.edu.cn}
}

\begin{document}
\maketitle
\begin{abstract}
    Most publicly accessible remote sensing data suffer from low resolution, limiting their practical applications. To address this, we propose a diffusion model guided by neural operators (NO) for continuous remote sensing image super-resolution (NeurOp-Diff). Neural operators are used to learn resolution representations at arbitrary scales, encoding low-resolution (LR) images into high-dimensional features, which are then used as prior conditions to guide the diffusion model for denoising. This effectively addresses the artifacts and excessive smoothing issues present in existing super-resolution (SR) methods, enabling the generation of high-quality, continuous super-resolution images. Specifically, we adjust the super-resolution scale by a scaling factor \( s \), allowing the model to adapt to different super-resolution magnifications. Furthermore, experiments on multiple datasets demonstrate the effectiveness of NeurOp-Diff. Our code is available at \href{https://github.com/zerono000/NeurOp-Diff}{https://github.com/zerono000/NeurOp-Diff}.
\end{abstract}    
\section{Introduction}

\begin{figure}
\begin{center}
\includegraphics[width=1\linewidth]{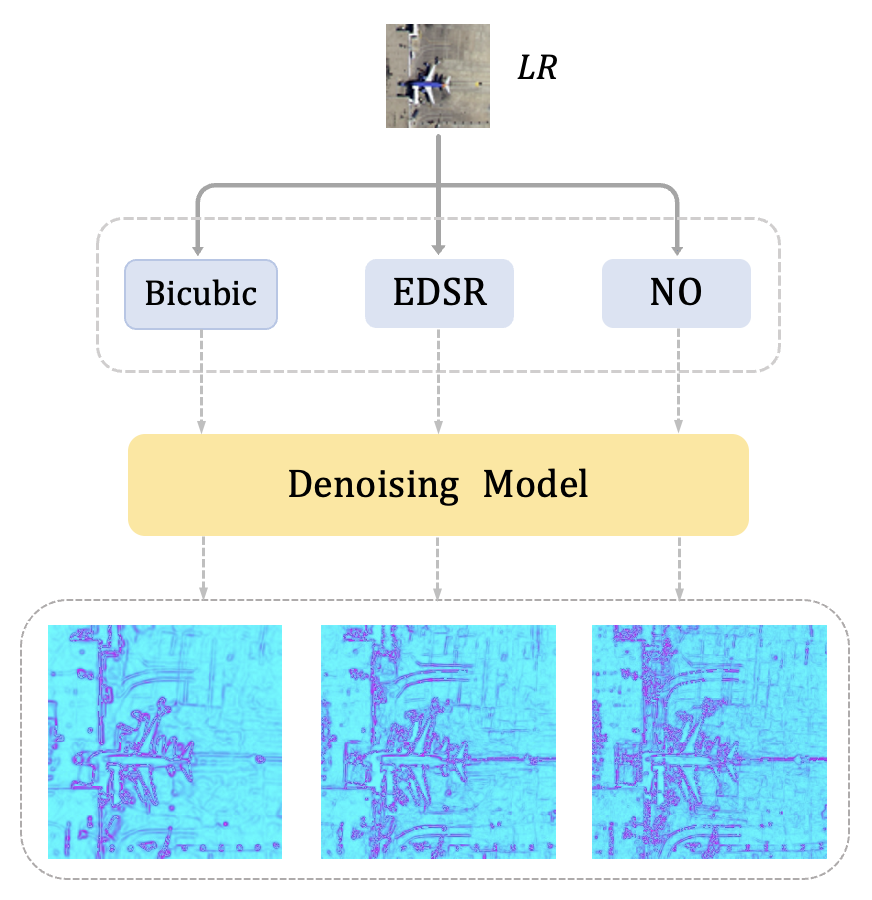}
\end{center}
   \caption{The figure illustrates the latent representations learned by the diffusion model after processing the low-resolution (LR) image using different methods, including Bicubic interpolation, EDSR, and Neural Operator (NO).}
\label{fig:1}
\end{figure}

Remote sensing super-resolution (RSSR) is a key technology in remote sensing image processing \cite{zhu2017deep}, aiming to restore low-resolution remote sensing (LRRS) images to high resolution remote sensing (HRRS) images, thus enhancing spatial details in remote sensing imagery \cite{wang2022comprehensive}. Despite advances in remote sensing technology that have greatly improved the spatial resolution of modern satellites, challenges related to resolution limitations persist \cite{bashir2021comprehensive}. On the other hand, different application scenarios, such as urban planning and agricultural monitoring, have varying requirements for image details and coverage. Continuous super-resolution can dynamically adjust the magnification factor according to specific needs \cite{chen2021learning, wei2023super}, providing greater adaptability. Moreover, remote sensing data comes from diverse sources with varying resolutions. Continuous super-resolution helps integrate data of different resolutions, improving the efficiency and quality of data fusion.

Early regression-based methods performed image super-resolution by establishing a mapping between low-resolution (LR) and high-resolution (HR) images. These approaches often relied on deep convolutional neural networks (CNNs) \cite{Dong2015, Dong2016, Kim2016a, Lim2017, vahdat2020nvae, Zhang2018}. The advantages of these methods include stable training, simple and easy-to-optimize loss functions, typically using mean squared error (MSE) to minimize pixel-wise differences between the reconstructed and ground-truth images. However, regression-based methods often result in overly smooth outputs that lack detail and struggle to recover realistic high-frequency information. On the other hand, variational autoencoders (VAEs) \cite{Kim2016b, Kingma2013} and generative adversarial networks (GANs) \cite{goodfellow2014generative, Karras2018,  karras2019style, ledig2017photo, wang2018esrgan,  dong2021rrsgan} have demonstrated exceptional image generation capabilities. GANs, in particular, introduce an adversarial learning mechanism to produce more realistic high-resolution images. Compared to regression-based methods, GANs-based approaches better preserve image details, generating sharper and more natural results. However, the adversarial training process is unstable, requiring careful hyperparameter tuning and often encountering mode collapse issues. Moreover, GANs generated images may suffer from artifacts, leading to distortions in specific areas of the reconstructed images \cite{goodfellow2014generative}.

Recently, diffusion models (DMs) \cite{ho2020denoising, sohl2015deep} have gained attention for their strong performance in generating high-quality images. Compared to GANs, diffusion models exhibit greater stability during training and can generate more complex and realistic image structures. In the context of image super-resolution, methods like SR3 \cite{saharia2022image} iteratively refine high-resolution images by progressively denoising from noisy inputs, achieving excellent reconstruction results. However, such methods are constrained by fixed scaling factors \cite{dai2019second, hsu2024drct, zhou2023learning}, requiring separate models for different magnification scales, which increases computational cost. Furthermore, due to the limited integration of additional prior knowledge, their performance still has room for improvement.

In this paper, we propose the diffusion model using neural operators as conditional guidance (NeurOp-Diff), for continuous super-resolution of remote sensing images. By leveraging the neural operator's ability to learn mappings between infinite-dimensional function spaces, we achieve continuous super-resolution for remote sensing images. The LRRS images processed by the neural operator serve as additional priors for the model's denoising process. Unlike previous works that used bicubic upsampled LR images as prior knowledge, the neural operator dynamically updates latent bases through a multilayer Galerkin-type attention mechanism \cite{wei2023super}. This process helps to recover certain high-frequency details, providing priors that enable the generation of more realistic high-resolution images \cite{rombach2022high} (see Figure \ref{fig:1}). Experimental results demonstrate that the images generated by NeurOp-Diff surpass those produced by previous methods, both in quality and detail.

The contributions of this paper are summarized as follows:
\begin{itemize}
    \item We propose a unified frame-work that integrates neural operators with diffusion models to remote sensing image super-resolution for the first time.
    \item We combined the high-frequency-rich continuous prior with the diffusion model, enabling the reverse sampling process to better recover the detailed features and texture information from the original data.
    \item We conducted extensive experiments on three remote sensing datasets, and compared with five advanced methods, our results showed greater competitiveness.
\end{itemize}

\begin{figure*}
\begin{center}
\includegraphics[width=0.9\linewidth]{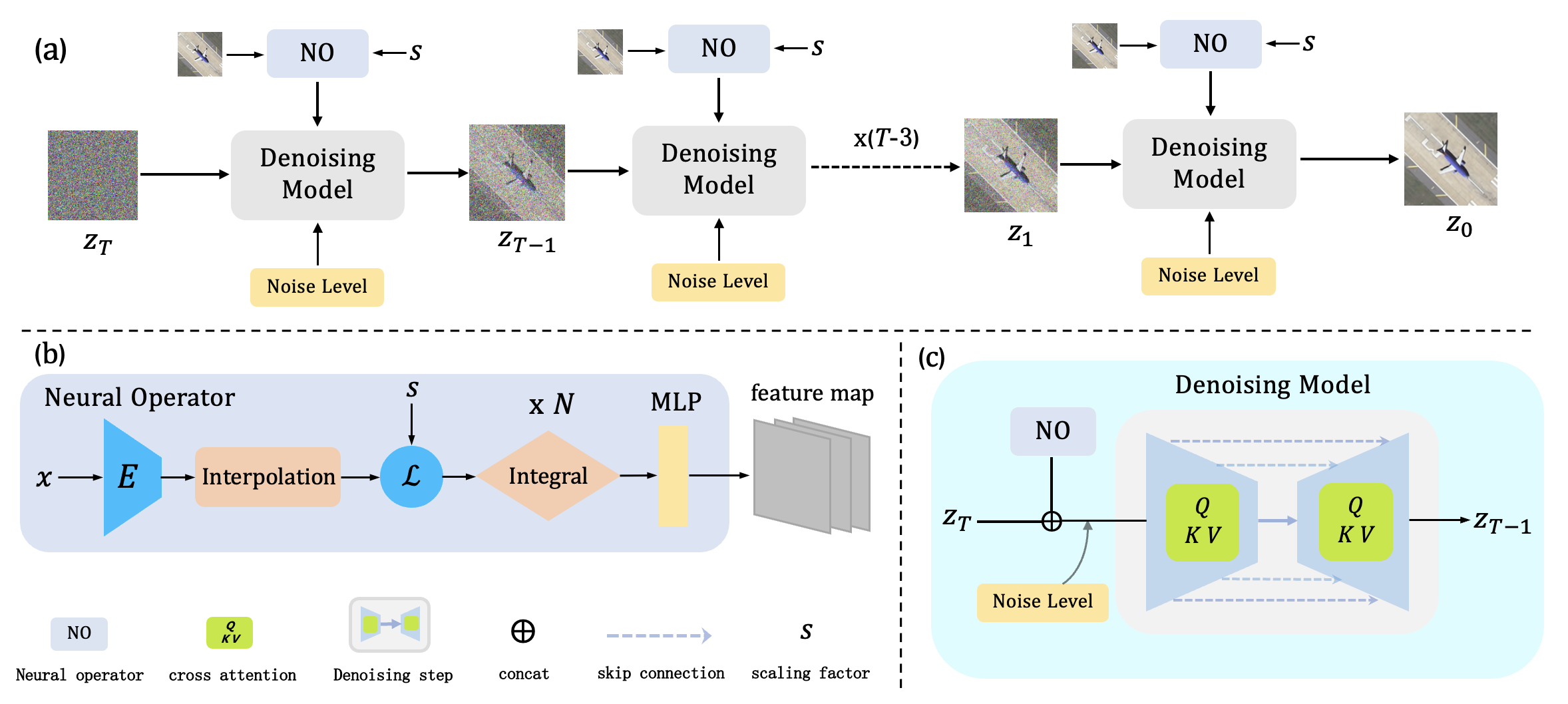}
\end{center}
   \caption{\textbf{(a)} The architecture of NeurOp-Diff. The noisy image is denoised through T iterations to generate a clear image. \( s \) is the scaling factor. \textbf{(b)} Neural operator architecture. The LR image is first encoded into high-dimensional features through the encoder (\textit{E}), interpolation, and lifting functions ($\mathcal{L}$), which linearly transform low-resolution pixel values. Then, it passes through a kernel integral composed of Galerkin-type attention to produce the output features. Finally, MPL is applied for channel transformation. \textbf{(c)} The integration of neural operators and diffusion models. Here, \( Q \), \( K \), and \( V \) represent the components of the attention mechanism.}
\label{fig:3}
\end{figure*}
\section{Relate Work}

\textbf{Remote Sensing Super-Resolution Generation Models.} Deep convolutional neural networks (CNNs) \cite{Dong2015, Dong2016, Kim2016a, Lim2017, vahdat2020nvae, Zhang2018} were among the earliest and most prominent methods for natural image super-resolution and have since been adapted to remote sensing image super-resolution (SR) due to their impressive performance. For example, LGCNet \cite{wu2023} uses a multi-branch structure to learn both local details and global environmental priors of remote sensing images, while RDBPN \cite{zhang2018residual} employs a residual dense back-projection network to achieve large-scale super-resolution reconstruction. With the successful application of Transformers in computer vision \cite{chen2021pre, dosovitskiy2020image, guo2022image, liu2021swin, vaswani2017attention, wang2022uformer}, Transformer-based super-resolution methods have started to gain significant attention. Although the SwinIR \cite{liang2021swinir} model has shown excellent image reconstruction performance, it tends to struggle when it comes to restoring high-frequency details and complex textures. Simultaneously, GANs \cite{goodfellow2014generative, Karras2018,  karras2019style, ledig2017photo,  dong2021rrsgan, wang2018esrgan} have shown strong potential for restoring image details. For example, EEGAN \cite{Ji2019} employs a dense edge-enhancement network to reduce noise and improve edge contours. However, GANs are known for their unstable training process, which can lead to model collapse.

In recent years, denoising diffusion models (DDMs) \cite{ho2020denoising, sohl2015deep, song2020denoising} have emerged as a leading approach in the image generation domain. The denoising diffusion probabilistic model (DDPM) \cite{ho2020denoising} introduced a new paradigm for image generation. Subsequent models such as Stable Diffusion \cite{rombach2022high} showcased the powerful capabilities of diffusion models in text-to-image generation, garnering significant attention. In specific tasks like image super-resolution, generative models have demonstrated unique advantages. For example, SR3 \cite{saharia2022image} achieves impressive super-resolution results by iterative refinement. However, its performance remains limited by insufficient prior knowledge. 
Recent works like Ref-Diff \cite{dong2024building} and SGDM \cite{wang2025semantic} incorporate semantic priors—land cover changes and vector maps—into conditional diffusion or generative models, enhancing fidelity and realism in large-scale remote sensing SR. However, they are still limited to fixed upscaling factors. To address this, we introduce neural operators based on SR3, enabling not only continuous super-resolution but also providing richer prior knowledge to enhance the denoising performance of the diffusion model.

\textbf{Neural Operators.} Neural operators are an emerging deep learning model capable of effectively learning and approximating complex nonlinear mappings while preserving the physical invariance and geometric structure of problems. This makes them particularly well-suited for solving complex partial differential equations \cite{operator1, operator2, operator3, operator4}. Unlike traditional neural networks, which typically learn mappings between finite-dimensional spaces, neural operators are designed to learn mappings from function space to function space, unconstrained by spatial dimensions. They can handle inputs of varying resolutions and sizes, producing corresponding functional outputs.

In recent years, neural operators have been applied to super-resolution tasks with promising results. For example, SRNO \cite{wei2023super} employs a Galerkin-type attention mechanism and a multilayer architecture to treat the mapping from low-resolution to high-resolution images as a transformation in continuous function space, enabling fast, efficient, and dynamically adaptive super-resolution reconstruction. However, such methods often exhibit over-smoothing when restoring high-frequency details and textures in images. To address this, we integrate denoising diffusion models into our work. Diffusion models are capable of generating high-quality images with rich details and natural textures. By combining diffusion models with neural operators, our approach retains the advantage of neural operators in achieving continuous super-resolution while mitigating the issue of over-smoothing in image reconstruction.

\section{Method}

Given a set of remote sensing image datasets $D = \{(x^{(i)}, y^{(i)}, z^{(i)})\}_{i=1}^N$ and a control output resolution factor $s$, the datasets satisfy a specific joint probability distribution $p(X, Y, Z)$, where $x^{(i)}$ denotes the low-resolution source image, $y^{(i)}$ denotes the degraded high-resolution image obtained through the network, and $z^{(i)}$ denotes the ground-truth high-resolution image. Using the approximation of the conditional distribution $p(z|y)$ \cite{wu2023hsr}, we aim to map low-resolution images to their corresponding high-resolution counterparts.

NeurOp-Diff first initializes $z^{(i)}$ as pure noise $z_T^{(i)} \sim \mathcal{N}(0, I)$, and then refines it iteratively $T$ times through the learned conditional transition distribution $p_\theta(z_{t-1}^{(i)}\mid z_t^{(i)}, y^{(i)})$ to generate high-resolution images $z_0^{(i)} \sim p(z\mid y)$. Here, $\theta$ denotes the learnable parameters of the neural network used to model the denoising process. The forward process defines how to progressively add noise to the initial $z_0^{(i)}$, generating the intermediate state sequence $(z_T^{(i)}, z_{T-1}^{(i)}, \ldots, z_0^{(i)})$ via a fixed Markov transition probability $q(z_t\mid z_{t-1})$. Our goal is to reverse this diffusion process using the reverse chain based on $y^{(i)}$, achieving the task of recovering the signal from noisy data. The reverse chain employs a denoising model built upon a U-Net architecture, taking as input low-resolution images processed by neural operators and noisy target images. Details of the model are as follows (see Figure \ref{fig:3}).

\subsection{Conditional Denoising Diffusion Model}

\subsubsection{Forward Process}
We first define a forward Markov diffusion process \cite{ho2020denoising} $q$, which progressively adds noise to the initial image $z_0$:
\begin{equation}
q(z_{1:T} \mid z_0) = \prod_{t=1}^{T} q(z_t \mid z_{t-1}).
\end{equation}

The transition probability at each step is expressed as:
\begin{equation}
q(z_t \mid z_{t-1}) = \mathcal{N}(z_t; \sqrt{\alpha_t} z_{t-1}, (1 - \alpha_t)I),
\end{equation}
where $\alpha_t$ is a hyperparameter controlling the noise level, satisfying $0 < \alpha_t < 1$. By marginalizing over intermediate steps, we can derive the boundary distribution of $z_t$ given $z_0$:
\begin{equation}
q(z_t \mid z_0) = \mathcal{N}(z_t; \sqrt{\gamma_t} z_0, (1 - \gamma_t)I),
\end{equation}
where $\gamma_t = \prod_{i=1}^t \alpha_i$. Using the properties of Gaussian distributions, the posterior distribution $q(z_{t-1} \mid z_0, z_t)$ can be derived as \cite{saharia2022image}:
\begin{equation}
q(z_{t-1} \mid z_0, z_t) = \mathcal{N}(z_{t-1}; \mu, \sigma^2 I),
\end{equation}
with:
\begin{equation}
\begin{gathered}
\mu = \frac{\sqrt{\gamma_{t-1}(1 - \alpha_t)}}{1 - \gamma_t} z_0 + \frac{\sqrt{\alpha_t}(1 - \gamma_{t-1})}{1 - \gamma_t} z_t, \\
\sigma^2 = \frac{(1 - \gamma_{t-1})(1 - \alpha_t)}{1 - \gamma_t}.
\end{gathered}
\end{equation}

\subsubsection{Reverse Process}

The inference stage is a reverse Markov process, which is the opposite of the forward diffusion process. It starts from pure Gaussian noise $z_T$ and progressively optimizes to generate $z_0$. The process is defined as:
\begin{equation}
\begin{gathered}
p_\theta(z_{0:T} \mid y) = p(z_T) \prod_{t=1}^T p_\theta(z_{t-1} \mid z_t, y), \\
p(z_T) = \mathcal{N}(z_T; 0, I), \\
p_\theta(z_{t-1} \mid z_t, y) = \mathcal{N}(z_{t-1}; \mu_\theta(y, z_t, \gamma_t), \sigma_t^2 I).
\label{eq:reverse_z1}
\end{gathered}
\end{equation}

By learning the conditional distribution of the forward diffusion process $p_\theta(z_{t-1} \mid z_t, y)$, the inference process can approximate the true posterior distribution \cite{sohl2015deep}, the final optimization step of the model is \cite{saharia2022image}:
\begin{equation}
z_{t-1} = \frac{1}{\sqrt{\gamma_t}} \left( z_t - \frac{1 - \alpha_t}{\sqrt{1 - \gamma_t}} f_\theta(y, z_t, \gamma_t) \right) + \sqrt{1 - \alpha_t} \epsilon_t,
\label{eq:reverse_z2}
\end{equation}
where $\epsilon_t \sim \mathcal{N}(0, I)$, and $f_\theta$ represents the noise prediction component.

\subsection{Neural Operator}
Neural operators map input functions to their latent representations and use kernel integration to capture global correlations in the function space. Specifically, we define a kernel function 
$K_t: \mathbb{R}^{d_r + d_r} \to \mathbb{R}^{d_r \times d_r}$, which operates on the hidden representations $\phi(\xi)$ and $\phi(\eta)$, rather than relying on spatial variables $(\xi, \eta)$ \cite{wei2023super}. Here, $d_r$ denotes the feature dimension, and $K_t$ represents the kernel generation function at the $t$-th layer. To efficiently explore the hidden representations $\phi(\xi)$, we will use multiple test functions to define these distributions, similar to single-head self-attention \cite{Kovachki2021}. Let $\phi_i = \phi(\xi_i) \in \mathbb{R}^{d_r}$, for $i = 1, \ldots, n_{h_f}$, where $n_{h_f}$ denotes the number of pixels on the high-resolution grid.

The kernel integral operator $\mathcal{K}(\phi)$ can be approximated using Monte Carlo methods, as follows:
\begin{equation}
\begin{gathered}
\mathcal{K}(\phi)(\xi) = \int_\Omega K(\phi(\xi), \phi(\eta)) \phi(\eta) \, d\eta \\
\approx \sum_{i=1}^{n_f} K(\phi(\xi), \phi_i) \phi_i, \quad \forall \xi \in \Omega_{h_f}
\end{gathered}
\end{equation}
where the kernel function $K(\phi(\xi), \phi_i)$ is computed as:
\begin{equation}
K(\phi(\xi), \phi_i) = 
\frac{\exp\left(\frac{\langle W_q \phi(\xi), W_k \phi_i \rangle}{\sqrt{d_r}}\right)}
{\sum_{j=1}^{n_f} \exp\left(\frac{\langle W_q \phi_j, W_k \phi_i \rangle}{\sqrt{d_r}}\right)} W_v
\end{equation}
Here, $W_q$, $W_k$, and $W_v$ correspond to the query, key, and value matrices. Using this formulation, the kernel function can represent a combination of trainable functions.

The advantages of neural operators lie in their ability to handle data with different resolutions while capturing global relationships in the function space through inner product operations \cite{wei2023super}. For example, a function $f$ can be represented as a distribution function instead of being evaluated at individual coordinate points. We use query, key, and value functions $q(\xi)$, $k(\xi)$, and $v(\xi)$ to express global correlations, and the kernel integration operation can be represented as $(j = 1, \ldots, d_r)$:
\begin{equation}
((\mathcal{K}(\phi))(\xi))_j = \sum_{l=1}^{d_r} \langle k_l, v_j \rangle q_l(\xi), \quad \forall \xi \in \Omega_{h_f}
\end{equation}

In the framework of neural operators, the latent representation $\phi$ is obtained through a linear combination of basis functions. To achieve efficient computation in high-resolution scenarios, Galerkin-type attention \cite{cao2021choose} is introduced. This mechanism achieves an approximately optimal solution through the linear combination of $Q$, $K$, and $V$:
\begin{equation}
\phi = Q(\tilde{K}^\top \tilde{V}) / n_f,
\end{equation}
where $\tilde{K}$ and $\tilde{V}$ are normalized keys and values. This method not only reduces computational complexity but also effectively addresses discontinuities at feature boundaries. To further enhance expressive power, neural operators iteratively update representations through a feed-forward network (FFN $\mathcal{O} : \mathbb{R}^{d_r} \to \mathbb{R}^{d_r}$). The iterative process is described as:
\begin{equation}
\phi_{t+1}(\xi) = \phi_t(\xi) + \mathcal{O}\left((K_t(\phi_t))(\xi) + \phi_t(\xi)\right).
\end{equation}

This iterative mechanism \cite{wei2023super} enables neural operators to continuously enrich their representational capabilities, improving performance in ultra-high-resolution tasks.

\subsection{Neural Operator with Conditional Diffusion}

Although the Conditional Denoising Diffusion Model excels in high-fidelity reconstruction of target images, with strong denoising capabilities and high-quality generation performance, its input priors are typically derived from simple upsampling operations (e.g., bicubic interpolation), which have limited expressive power and are constrained by fixed scaling factors. In contrast, the Neural Operator can learn mappings between infinite-dimensional function spaces and extract latent continuous high-frequency information, effectively addressing the limitations of input prior expressiveness and fixed resolution. Therefore, we incorporate the Neural Operator into the Conditional Denoising Diffusion Model to enhance the expressive power of input priors, thereby improving the quality of generated images and enabling more flexible continuous super-resolution. Specifically, we first employ the Neural Operator \( \mathcal{F}_\theta \) to process the low-resolution remote sensing image, obtaining a latent continuous prior representation:

\begin{equation}
    y = \mathcal{F}_\theta(x),
\end{equation}

where \( y \) encapsulates richer and more dynamic high-frequency information compared to traditional upsampling priors. Subsequently, during the reverse process of the diffusion model (as detailed in Equations~(\ref{eq:reverse_z1}) and~(\ref{eq:reverse_z2})), this Neural Operator prior \( y \) is incorporated as conditional information into the noise prediction network \( \mathcal{G}_\theta \), thereby refining the sampling distribution at each diffusion inversion step.

In our model, given the intermediate state \( z_t \) in the diffusion process and the prior derived from the neural operator \( y \), the reverse update equation is expressed as:

\begin{equation}
    z_{t-1} = \frac{1}{\sqrt{\gamma_t}} \left( z_t - \frac{1 - \alpha_t}{\sqrt{1 - \gamma_t}} \mathcal{G}_\theta(y, z_t, \gamma_t) \right) + \sqrt{1 - \alpha_t} \, \epsilon_t
\end{equation}

In this equation, \( \mathcal{G}_\theta(y, z_t, \gamma_t) \) serves not only as the noise prediction component but also carries the latent function space prior provided by the Neural Operator. This integration allows for a more flexible and expressive guidance of the diffusion process updates. Through multi-layer Galerkin-type attention mechanisms and dynamically updated latent bases, the neural operator captures complex features and high-frequency information at a global scale of the input image. When this information is incorporated as conditional guidance into the diffusion process, the model can progressively eliminate noise and converge toward more realistic and higher-resolution remote sensing image reconstructions during iterative sampling.

\subsection{Network details}
NeurOp-Diff network architecture is illustrated in the figure \ref{fig:3}. The neural operator utilizes a scaling factor $s \sim \mathcal{U}(1, M]$ to ensure the model is effectively applicable to multiple scales. In this experiment, we set $M = 8$. For the feature encoder $E$, we reference the EDSR-baseline \cite{Lim2017} by removing the last sampling layer. Regarding the noise level $\gamma$, we follow a uniform sampling method. First, the time step $t \sim \{0, T\}$ is sampled uniformly, followed by sampling $\gamma$ from $\mathcal{U}(\gamma_{t-1}, \gamma_t)$. This process is expressed as: $\gamma \sim \frac{1}{T} \sum_{t=1}^T \mathcal{U}(\gamma_{t-1}, \gamma_t)$ \cite{saharia2022image}.

\begin{figure*}
\begin{center}
\includegraphics[width=\textwidth]{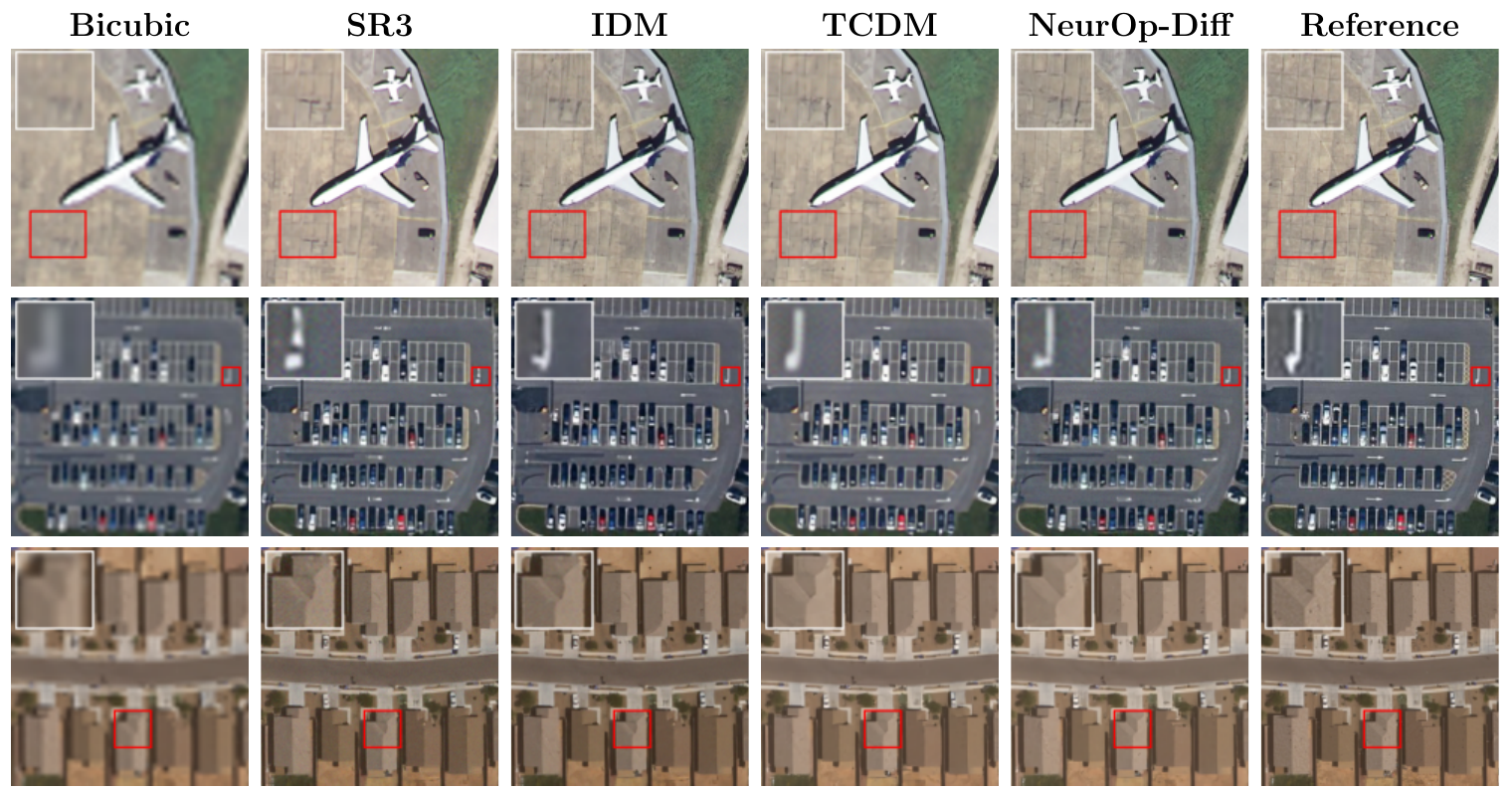}
\end{center}
   \caption{A visualization quality comparison between NeurOp-Diff and generative super-resolution (SR) models on UCMerced and AID. For each image, we also display finer details of the magnified regions.}
\label{fig:4}
\end{figure*}
\section{Experiments}

\subsection{Training}
\paragraph{Experimental Setup.} In this experiment, we used three remote sensing datasets: UCMerced \cite{yang2010bag}, AID \cite{xia2017aid} and RSSCN7 \cite{zou2015}. A portion of each dataset was used as the training set, while the remaining portion was used for evaluation. For AID \cite{xia2017aid} and RSSCN7 \cite{zou2015}, which do not match the target resolution, we adjusted the data using center cropping \cite{brock2018large} and downsampling. We compared NeurOp-Diff with various representative super-resolution methods, including both diffusion-based generative approaches and regression-based continuous super-resolution methods. To evaluate the effectiveness of our continuous super-resolution method, we also tested magnification factors beyond the training distribution when compared with SRNO \cite{wei2023super}. During the training phase, the scaling factors were uniformly sampled within a continuous range from ×1 to ×8. In the testing phase, the model was further evaluated on images with out-of-distribution higher scaling factors, specifically from ×9 to ×10. The experimental results are primarily evaluated using PSNR (Peak Signal-to-Noise Ratio) and SSIM (Structural Similarity Index) \cite{wang2004image}.

\paragraph{Training Details.} Following the referenced approach \cite{wei2023super}, we first pre-trained the neural operator component for 500 epochs with a batch size of 64. Then, we froze the weights of the neural operator and trained the entire NeurOp-Diff model. The batch size was set to 10, and approximately 1M iterations were performed. The Adam optimizer \cite{kingma2014adam} was used, with an initial learning rate of $1 \times 10^{-4}$ and a minimum learning rate of $2 \times 10^{-6}$. For the first 0.1 M iterations, a constant learning rate was maintained, followed by a learning rate decay strategy. Dropout was set to 0.2 during training. During training, the scaling factor $s \sim \mathcal{U}(1, M]$ was sampled from a uniform distribution to cover multiple levels of magnification. For super-resolution tasks, a larger $T$ generally yields better results. Therefore, in this experiment, we set $T = 2000$. And we chose the $L_1$ norm as the loss function. In the inference phase, we used a uniform sampling strategy \cite{song2020denoising} to accelerate inference. Experiments showed that setting the number of noise diffusion steps to 50 produced good inference results. The experiments were conducted on an NVIDIA RTX 4090 GPU.

\begin{table}[htbp]
\centering
\begin{tabular}{p{1.3cm}p{2cm}p{0.6cm}p{0.7cm}p{0.6cm}p{0.7cm}}
\toprule
\textbf{Dataset} & \textbf{Methods} & 
\multicolumn{2}{c}{\textbf{x4}} & 
\multicolumn{2}{c}{\textbf{x8}} \\
\cmidrule(lr){3-4} \cmidrule(lr){5-6} 
& & \textbf{PSNR} & \textbf{SSIM} & \textbf{PSNR} & \textbf{SSIM} \\

\midrule
\multirow{4}{*}{UCMerced} 
& SR3 \cite{saharia2022image}         & 27.15 & 0.7587 & 23.12 & 0.5846 \\
& IDM \cite{gao2023implicit}        & 27.66 & 0.7606 & 23.68 & 0.6131 \\
& TCDM \cite{zhang2024}       & 27.83 & 0.7679 & 24.17 & 0.6422 \\
& \textbf{NeurOp-Diff} & \textbf{28.14} & \textbf{0.7761} & \textbf{24.51} & \textbf{0.6502} \\
\midrule
\multirow{4}{*}{AID} 
& SR3 \cite{saharia2022image}       & 26.24 & 0.6709 & 22.54 & 0.5231 \\
& IDM \cite{gao2023implicit}       & 27.03 & 0.6778 & 22.90 & 0.5487 \\
& TCDM \cite{zhang2024}      & 27.53 & 0.6840 & 23.37 & 0.5584 \\
& \textbf{NeurOp-Diff} & \textbf{27.57} & \textbf{0.6845} & \textbf{23.45} & \textbf{0.5599} \\
\midrule
\multirow{4}{*}{RSSCN7} 
& SR3 \cite{saharia2022image}         & 26.22 & 0.5835 & 22.96 & 0.5329\\
& IDM \cite{gao2023implicit}         & 27.18 & 0.6235 & 23.37 & 0.5603 \\
& TCDM \cite{zhang2024}    & 27.71 & 0.6489 & 23.69 & 0.5873 \\
& \textbf{NeurOp-Diff} & \textbf{27.74} & \textbf{0.6509} & \textbf{23.73} & \textbf{0.5889} \\
\bottomrule
\end{tabular}
\medskip
\caption{Quantitative comparison of various generative super-resolution ($64 \times 64 \rightarrow 256 \times 256$, $32 \times 32 \rightarrow 256 \times 256$) methods on UCMerced, AID, and RSSCN7 in terms of average PSNR and SSIM. The best performance are bolded.}
\label{tab:1}
\end{table}

\subsection{Comparison of Generative SR}

\subsubsection{Qualitative Comparison}
Figure \ref{fig:4} presents the super-resolution results of $64 \times 64 \to 256 \times 256$ for the UCMerced and AID remote sensing datasets. Compared with various state-of-the-art (SOTA) generative methods, NeurOp-Diff demonstrates excellent reconstruction capability. Although all methods can generally produce realistic super-resolution images, the magnified regions marked with rectangles reveal that NeurOp-Diff demonstrates superior reconstruction performance in recovering high-frequency components such as road textures and shadows, compared to other generative methods. This is attributed to the prior knowledge provided by the NO.

\subsubsection{Quantitative Comparison}
Table \ref{tab:1} presents the PSNR and SSIM results of various models for 4x SR and 8x SR. Compared to other generative models, NeurOp-Diff demonstrates superior performance in both pixel-level accuracy (PSNR) and visual perceptual quality (SSIM).

\subsubsection{Runtime Efficiency Comparison}
Table \ref{tab:running_time_comparison} shows a comparison of runtime between NeurOp-Diff and SR3, IDM under different inference steps. Although the runtime of all methods increases with the number of inference steps, our method maintains the fastest execution speed across all tested configurations.

\begin{table}[htbp]
  \centering
  \renewcommand{\arraystretch}{1}
    \begin{tabular}{lcccc}
    \toprule
    Inference Steps & SR3 \cite{saharia2022image} & IDM \cite{gao2023implicit} & NeurOp-Diff \\
    \midrule
    T=30 & 1.8 & 1.47 & \textbf{1.12} \\
    T=50 & 2.16 & 1.79 & \textbf{1.52} \\
    T=100 & 3.23 & 2.87 & \textbf{2.61} \\
    \bottomrule
  \end{tabular}
  \smallskip
  \caption{A comparison of the inference time (in seconds) required by different methods to process a single image under varying numbers of inference steps (T = 30, 50, 100).}
  \label{tab:running_time_comparison}
\end{table}

\begin{figure}
\begin{center}
\includegraphics[width=1\linewidth]{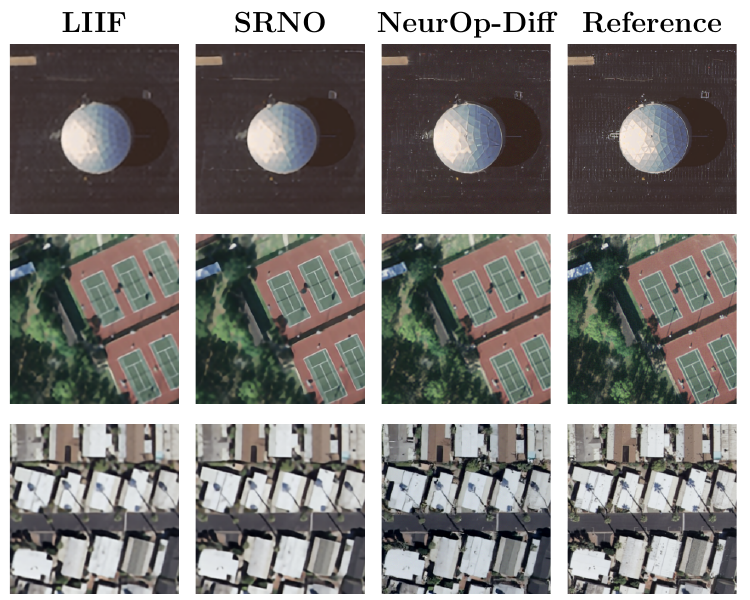}
\end{center}
   \caption{Qualitative comparison on 4× SR on UCMerced. Compared to regression-based methods, NeurOp-Diff is able to reconstruct more details and produce sharper textures.}
\label{fig:6}
\end{figure}

\begin{table*}[htbp]
\centering
\renewcommand{\arraystretch}{1.1}
\caption{Quantitative comparison of continuous super-resolution (SR) results on UCMerced (PSNR/SSIM/LPIPS).}
\medskip
\begin{tabular}{llccccccc}
    \toprule
    \multirow{2}{*}{Method} & \multirow{2}{*}{Metric} & \multicolumn{5}{c}{in-distribution} & \multicolumn{2}{c}{out-of-distribution} \\
    \cmidrule(lr){3-7} \cmidrule(lr){8-9}
     & & $2\times$ & $3.1\times$ & $5.8\times$ & $7\times$ & $8\times$ & $9\times$ & $10\times$ \\
    \midrule
    \multirow{3}{*}{SRNO \cite{wei2023super}} 
    & PSNR & \textbf{34.32} & \textbf{31.05} & \textbf{27.19} & \textbf{25.73} & \textbf{24.67} & \textbf{24.32} & \textbf{23.92} \\
    & SSIM & 0.786 & 0.746 & 0.681 & 0.649 & 0.623 & 0.607 & 0.592 \\
    & LPIPS & 0.118 & 0.142	& 0.194 & 0.207 & 0.215 & 0.221 & 0.228 \\
    \midrule
    \multirow{3}{*}{IDM \cite{gao2023implicit}}
    & PSNR & 33.54 & 30.32 & 26.37 & 25.02 & 23.68 & 23.34 & 23.05 \\
    & SSIM & 0.811 & 0.781 & 0.692 & 0.646 & 0.613 & 0.601 & 0.587 \\
    & LPIPS & 0.103	& 0.119 & 0.187 & 0.208 & 0.219 & 0.225 & 0.230 \\
    \midrule
    \multirow{3}{*}{NeurOp-Diff}
    & PSNR & 33.93 & 30.76 & 26.89 & 25.50 & 24.51 & 24.13 & 23.88 \\
    & SSIM & \textbf{0.824} & \textbf{0.792} & \textbf{0.717} & \textbf{0.673} & \textbf{0.650} & \textbf{0.641} & \textbf{0.633} \\
    & LPIPS & \textbf{0.098} & \textbf{0.112} & \textbf{0.159} & \textbf{0.188} & \textbf{0.195} & \textbf{0.206} & \textbf{0.212} \\
    \bottomrule
\end{tabular}
\label{tab:3}
\end{table*}

\begin{figure*}
\begin{center}
\includegraphics[width=1\linewidth]{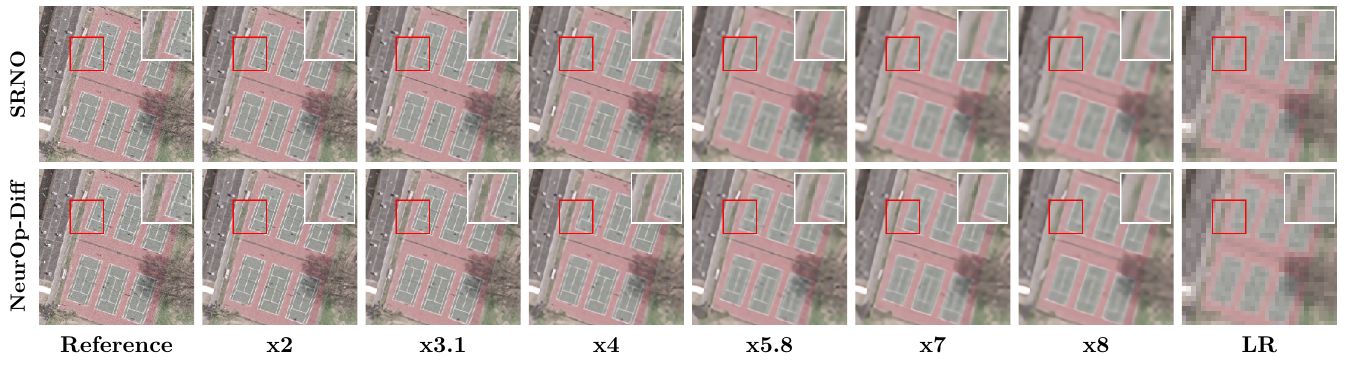}
\end{center}
   \caption{Visualization of continuous SR results on UCMerced. The resolution of the ground truth image is $256 \times 256$. We randomly selected multiple magnification factors within the range of (1, 8] for demonstration, additionally including two out-of-distribution magnification factors. (i.e., 5.8x, 8x, and 10x).}
\label{fig:7}
\end{figure*}

\subsection{Comparison of Continuous SR}
Table \ref{tab:2} presents the quantitative comparison of LIIF, SRNO, and NeurOp-Diff on UCMerced. Although LIIF and SRNO achieve higher PSNR, our NeurOp-Diff demonstrates better performance in terms of SSIM. Moreover, in subjective perceptual differences (LPIPS) \cite{zhang2018unreasonable}, the images generated by our model show smaller differences compared to the real images. As shown in Figure \ref{fig:6}, regression-based continuous super-resolution methods struggle to reconstruct high-frequency details and restore realistic textures, often producing overly smooth images. For instance, fine ground stripes are almost invisible in the images output by SRNO. Even when reconstructing more prominent stripes, the results appear somewhat blurry. In contrast, NeurOp-Diff demonstrates superior capability in restoring these details.

Furthermore, we randomly selected multiple magnification factors from the range (1, 8] and visualized images at these magnifications using both SRNO and NeurOp-Diff methods in Figure \ref{fig:7}, in order to validate the performance of NeurOp-Diff in the continuous super-resolution task. Specifically, the real images were first downsampled according to the selected magnification factors to obtain corresponding low-resolution images. Then, SRNO and NeurOp-Diff were used to reconstruct these low-resolution images, generating high-resolution output images, such as 2x SR ($128 \times 128 \to 256 \times 256$), 4x SR ($64 \times 64 \to 256 \times 256$) and 8x SR ($32 \times 32 \to 256 \times 256$). During the reconstruction process, the output image resolution was fixed and unchanged.

From a quantitative perspective (see Table \ref{tab:3}), although SRNO demonstrates slightly better PSNR values at lower magnification factors (e.g., 2x to 3.1x), NeurOp-Diff demonstrates superior performance in terms of SSIM metrics, which better aligns with human visual perception. As the magnification factor increases, the advantages of NeurOp-Diff become more apparent, particularly in challenging scenarios with higher magnification factors (e.g., 7x to 10x), where it consistently maintains higher SSIM values as well as lower LPIPS scores. Visually, NeurOp-Diff exhibits better performance in terms of the clarity of tennis court boundary lines and natural texture transitions in the surrounding environment. These advantages become more pronounced at higher magnification factors. This is primarily because NeurOp-Diff is better at capturing and reconstructing high-frequency information, allowing the generated images at high magnification factors to maintain high visual quality and detail richness. However, SRNO often struggles to fully recover complex texture details when handling larger magnifications, resulting in relatively lacking detail in the generated images.

\begin{table}[ht]
\centering
\begin{tabular}{lccc}
\toprule
\textbf{Method} & \textbf{PSNR$\uparrow$} & \textbf{SSIM$\uparrow$} & \textbf{LPIPS$\downarrow$} \\
\midrule
LIIF \cite{chen2021learning}  & 28.96 & 0.69 & 0.182  \\
SRNO \cite{wei2023super}  & \textbf{29.21} & 0.72 & 0.164  \\
\midrule
NeurOp-Diff         & 28.14 & \textbf{0.78} & \textbf{0.136} \\
\bottomrule
\end{tabular}
\caption{Quantitative comparison of NeurOp-Diff and regression methods for 4x SR on UCMerced.}
\label{tab:2}
\end{table}

\subsection{Ablation Studies}

To explore the impact of the LR conditional network, we conducted 4× super-resolution ablation experiments on remote sensing datasets. Specifically, we constructed comparison models by replacing the scale-adaptive conditional network in NeurOp-Diff with three different conditional mechanisms: (1) Concatenating the upsampled LR image directly with the ground truth image (SR3 \cite{saharia2022image}); (2) Concatenating LR features encoded by EDSR \cite{Lim2017} with the ground truth image; (3) Concatenating LR features encoded by the neural operator with the ground truth image. As shown in Figure \ref{fig:1}, directly using the upsampled LR image as a prior condition results in the least amount of detail recovery. Encoding LR features with EDSR recovers more details, but it still falls short compared to the scale-adaptive network encoded by the neural operator. The quantitative results in Table \ref{tab:comparison} further validate the effectiveness of the neural operator-based conditioning mechanism, as it achieves the best performance in both PSNR and SSIM. In comparison, it is evident that LR images encoded by the neural operator provide more prior knowledge.

\begin{table}[htbp]
\centering
\begin{tabular}{lccc}
\toprule
& Bicubic & EDSR & Neural Operator \\
\midrule
\textbf{PSNR$\uparrow$} & 27.15 & 27.42 & \textbf{28.14} \\
\textbf{SSIM$\uparrow$} & 0.7587 & 0.7602 & \textbf{0.7761} \\
\bottomrule
\end{tabular}
\caption{Comparison of different methods by PSNR and SSIM}
\label{tab:comparison}
\end{table}

\section{Conclusion}

This paper presented Continuous Remote Sensing Image Super-Resolution via Neural Operator Diffusion (NeurOp-Diff). The method uses the LR remote sensing image encoded by the neural operator as the prior condition for the denoising process of the diffusion model, and achieves continuous super-resolution through the neural operator's ability to map to infinite-dimensional function spaces. Experiments show that, compared to various advanced methods, our approach achieves superior results.

\section*{Acknowledgements}

This work was supported by the National Natural Science Foundation of China (No. 42306245),  in part by the National Key Research and Development Program of China (No. 2024YFF0617900), in part by the Shenzhen Outstanding Talents Training Fund, in part by the Shenzhen Major Science and Technology Project (No. KCXFZ20240903093000002), and in part by the Open Research Fund Program of MNR Key Laboratory for Geo-Environmental Monitoring of Great Bay Area (No. GEMLab-2023015).
{
    \small
    \bibliographystyle{ieeenat_fullname}
    \bibliography{main}

@String(ECCV= {Eur. Conf. Comput. Vis.})

@String(AAAI = {AAAI})

@String(ECCV  = {ECCV})

@inproceedings{operator1,
  author    = {Gaurav Gupta and Xiongye Xiao and Paul Bogdan},
  title     = {Multiwavelet-based operator learning for differential equations},
  booktitle = {Advances in Neural Information Processing Systems},
  volume    = {34},
  pages     = {24048--24062},
  year      = {2021}
}

@inproceedings{operator2,
  author    = {Rakhoon Hwang and Jae Yong Lee and Jin Young Shin and Hyung Ju Hwang},
  title     = {Solving pde-constrained control problems using operator learning},
  booktitle = {Proceedings of the AAAI Conference on Artificial Intelligence},
  volume    = {36},
  number    = {4},
  pages     = {4504--4512},
  year      = {2022}
}

@article{operator3,
  title={Fourier neural operator for parametric partial differential equations},
  author={Li, Zongyi and Kovachki, Nikola and Azizzadenesheli, Kamyar and Liu, Burigede and Bhattacharya, Kaushik and Stuart, Andrew and Anandkumar, Anima},
  journal={arXiv preprint arXiv:2010.08895},
  year={2020}
}

@article{operator4,
  author    = {Lu Lu and Pengzhan Jin and Guofei Pang and Zhongqiang Zhang and George Em Karniadakis},
  title     = {Learning nonlinear operators via DeepONet based on the universal approximation theorem of operators},
  journal   = {Nature Machine Intelligence},
  volume    = {3},
  number    = {3},
  pages     = {218--229},
  year      = {2021}
}

@inproceedings{Dong2015,
  author    = {C. Dong and C. C. Loy and K. He and X. Tang},
  title     = {Image super-resolution using deep convolutional networks},
  booktitle = {Proceedings of the IEEE Conference on Computer Vision and Pattern Recognition},
  pages     = {184--192},
  year      = {2015}
}

@inproceedings{Kim2016a,
  author    = {J. Kim and J. K. Lee and K. M. Lee},
  title     = {Deeply-recursive convolutional network for image super-resolution},
  booktitle = {Proceedings of the IEEE Conference on Computer Vision and Pattern Recognition},
  pages     = {1637--1645},
  year      = {2016}
}

@inproceedings{Dong2016,
  author    = {C. Dong and C. C. Loy and X. Tang},
  title     = {Accelerating the super-resolution convolutional neural network},
  booktitle = {Proceedings of the European Conference on Computer Vision},
  pages     = {391--407},
  year      = {2016}
}

@inproceedings{Kim2016b,
  author    = {J. Kim and J. K. Lee and K. M. Lee},
  title     = {Accurate image super-resolution using very deep convolutional networks},
  booktitle = {Proceedings of the IEEE Conference on Computer Vision and Pattern Recognition},
  pages     = {1646--1654},
  year      = {2016}
}

@inproceedings{Kingma2013,
  author    = {Diederik P. Kingma and Max Welling},
  title     = {Auto-Encoding Variational Bayes},
  booktitle = {Proceedings of the International Conference on Learning Representations},
  year      = {2013}
}

@article{vahdat2020nvae,
  title={NVAE: A deep hierarchical variational autoencoder},
  author={Vahdat, Arash and Kautz, Jan},
  journal={Advances in neural information processing systems},
  volume={33},
  pages={19667--19679},
  year={2020}
}

@inproceedings{Lim2017,
  author    = {B. Lim and S. Son and H. Kim and S. Nah and K. M. Lee},
  title     = {Enhanced deep residual networks for single image super-resolution},
  booktitle = {Proceedings of the IEEE Conference on Computer Vision and Pattern Recognition Workshops},
  pages     = {136--144},
  year      = {2017}
}

@inproceedings{Zhang2018,
  author    = {Y. Zhang and Y. Tian and Y. Kong and B. Zhong and Y. Fu},
  title     = {Residual dense network for image super-resolution},
  booktitle = {Proceedings of the IEEE Conference on Computer Vision and Pattern Recognition},
  pages     = {2472--2481},
  year      = {2018}
}

@article{goodfellow2014generative,
  title={Generative adversarial nets},
  author={Goodfellow, Ian and Pouget-Abadie, Jean and Mirza, Mehdi and Xu, Bing and Warde-Farley, David and Ozair, Sherjil and Courville, Aaron and Bengio, Yoshua},
  journal={Advances in neural information processing systems},
  volume={27},
  year={2014}
}

@inproceedings{Karras2018,
  author    = {Tero Karras and Timo Aila and Samuli Laine and Jaakko Lehtinen},
  title     = {Progressive growing of GANs for improved quality, stability, and variation},
  booktitle = {Proceedings of the International Conference on Learning Representations},
  year      = {2018}
}

@inproceedings{karras2019style,
  title={A style-based generator architecture for generative adversarial networks},
  author={Karras, Tero and Laine, Samuli and Aila, Timo},
  booktitle={Proceedings of the IEEE/CVF conference on computer vision and pattern recognition},
  pages={4401--4410},
  year={2019}
}

@inproceedings{ledig2017photo,
  title={Photo-realistic single image super-resolution using a generative adversarial network},
  author={Ledig, Christian and Theis, Lucas and Husz{\'a}r, Ferenc and Caballero, Jose and Cunningham, Andrew and Acosta, Alejandro and Aitken, Andrew and Tejani, Alykhan and Totz, Johannes and Wang, Zehan and others},
  booktitle={Proceedings of the IEEE conference on computer vision and pattern recognition},
  pages={4681--4690},
  year={2017}
}

@inproceedings{sohl2015deep,
  title={Deep unsupervised learning using nonequilibrium thermodynamics},
  author={Sohl-Dickstein, Jascha and Weiss, Eric and Maheswaranathan, Niru and Ganguli, Surya},
  booktitle={International conference on machine learning},
  pages={2256--2265},
  year={2015}
}

@article{ho2020denoising,
  title={Denoising diffusion probabilistic models},
  author={Ho, Jonathan and Jain, Ajay and Abbeel, Pieter},
  journal={Advances in neural information processing systems},
  volume={33},
  pages={6840--6851},
  year={2020}
}

@article{saharia2022image,
  title={Image super-resolution via iterative refinement},
  author={Saharia, Chitwan and Ho, Jonathan and Chan, William and Salimans, Tim and Fleet, David J and Norouzi, Mohammad},
  journal={IEEE transactions on pattern analysis and machine intelligence},
  volume={45},
  number={4},
  pages={4713--4726},
  year={2022}
}

@inproceedings{wei2023super,
  title={Super-resolution neural operator},
  author={Wei, Min and Zhang, Xuesong},
  booktitle={Proceedings of the IEEE/CVF Conference on Computer Vision and Pattern Recognition},
  pages={18247--18256},
  year={2023}
}

@inproceedings{rombach2022high,
  title={High-resolution image synthesis with latent diffusion models},
  author={Rombach, Robin and Blattmann, Andreas and Lorenz, Dominik and Esser, Patrick and Ommer, Bj{\"o}rn},
  booktitle={Proceedings of the IEEE/CVF conference on computer vision and pattern recognition},
  pages={10684--10695},
  year={2022}
}

@inproceedings{wu2023,
  author={Wu, Tzu-Han and Chen, Kuan-Wen},
  booktitle={2023 IEEE International Conference on Robotics and Automation}, 
  title={LGCNet: Feature Enhancement and Consistency Learning Based on Local and Global Coherence Network for Correspondence Selection}, 
  pages={6182-6188},
  year={2023}  
}

@inproceedings{zhang2018residual,
  title={Residual dense network for image super-resolution},
  author={Zhang, Yulun and Tian, Yapeng and Kong, Yu and Zhong, Bineng and Fu, Yun},
  booktitle={Proceedings of the IEEE conference on computer vision and pattern recognition},
  pages={2472--2481},
  year={2018}
}

@article{vaswani2017attention,
  title={Attention is all you need},
  author={Vaswani, A},
  journal={Advances in Neural Information Processing Systems},
  year={2017}
}

@article{dosovitskiy2020image,
  title={An image is worth 16x16 words: Transformers for image recognition at scale},
  author={Dosovitskiy, Alexey},
  journal={arXiv preprint arXiv:2010.11929},
  year={2020}
}

@inproceedings{liu2021swin,
  title={Swin transformer: Hierarchical vision transformer using shifted windows},
  author={Liu, Ze and Lin, Yutong and Cao, Yue and Hu, Han and Wei, Yixuan and Zhang, Zheng and Lin, Stephen and Guo, Baining},
  booktitle={Proceedings of the IEEE/CVF international conference on computer vision},
  pages={10012--10022},
  year={2021}
}

@inproceedings{chen2021pre,
  title={Pre-trained image processing transformer},
  author={Chen, Hanting and Wang, Yunhe and Guo, Tianyu and Xu, Chang and Deng, Yiping and Liu, Zhenhua and Ma, Siwei and Xu, Chunjing and Xu, Chao and Gao, Wen},
  booktitle={Proceedings of the IEEE/CVF conference on computer vision and pattern recognition},
  pages={12299--12310},
  year={2021}
}

@inproceedings{guo2022image,
  title={Image dehazing transformer with transmission-aware 3d position embedding},
  author={Guo, Chun-Le and Yan, Qixin and Anwar, Saeed and Cong, Runmin and Ren, Wenqi and Li, Chongyi},
  booktitle={Proceedings of the IEEE/CVF conference on computer vision and pattern recognition},
  pages={5812--5820},
  year={2022}
}

@inproceedings{wang2022uformer,
  title={Uformer: A general u-shaped transformer for image restoration},
  author={Wang, Zhendong and Cun, Xiaodong and Bao, Jianmin and Zhou, Wengang and Liu, Jianzhuang and Li, Houqiang},
  booktitle={Proceedings of the IEEE/CVF conference on computer vision and pattern recognition},
  pages={17683--17693},
  year={2022}
}

@inproceedings{liang2021swinir,
  title={Swinir: Image restoration using swin transformer},
  author={Liang, Jingyun and Cao, Jiezhang and Sun, Guolei and Zhang, Kai and Van Gool, Luc and Timofte, Radu},
  booktitle={Proceedings of the IEEE/CVF international conference on computer vision},
  pages={1833--1844},
  year={2021}
}

@article{Ji2019,
  author={Jiang, Kui and Wang, Zhongyuan and Yi, Peng and Wang, Guangcheng and Lu, Tao and Jiang, Junjun},
  journal={IEEE Transactions on Geoscience and Remote Sensing}, 
  title={Edge-Enhanced GAN for Remote Sensing Image Superresolution}, 
  year={2019},
  volume={57},
  number={8},
  pages={5799-5812},
}

@article{song2020denoising,
  title={Denoising diffusion implicit models},
  author={Song, Jiaming and Meng, Chenlin and Ermon, Stefano},
  journal={arXiv preprint arXiv:2010.02502},
  year={2020}
}

@article{Kovachki2021,
  author    = {Nikola Kovachki and Zongyi Li and Burigede Liu and Kamyar Azizzadenesheli and Kaushik Bhattacharya and Andrew Stuart and Anima Anandkumar},
  title     = {Neural Operator: Learning Maps Between Function Spaces},
  journal   = {arXiv preprint arXiv:2108.08481},
  year      = {2021}
}

@article{cao2021choose,
  title={Choose a transformer: Fourier or galerkin},
  author={Cao, Shuhao},
  journal={Advances in neural information processing systems},
  volume={34},
  pages={24924--24940},
  year={2021}
}

@inproceedings{yang2010bag,
  title={Bag-of-visual-words and spatial extensions for land-use classification},
  author={Yang, Yi and Newsam, Shawn},
  booktitle={Proceedings of the 18th SIGSPATIAL international conference on advances in geographic information systems},
  pages={270--279},
  year={2010}
}

@article{xia2017aid,
  title={AID: A benchmark data set for performance evaluation of aerial scene classification},
  author={Xia, Gui-Song and Hu, Jingwen and Hu, Fan and Shi, Baoguang and Bai, Xiang and Zhong, Yanfei and Zhang, Liangpei and Lu, Xiaoqiang},
  journal={IEEE Transactions on Geoscience and Remote Sensing},
  volume={55},
  number={7},
  pages={3965--3981},
  year={2017},
}

@article{zou2015,
  author={Zou, Qin and Ni, Lihao and Zhang, Tong and Wang, Qian},
  journal={IEEE Geoscience and Remote Sensing Letters}, 
  title={Deep Learning Based Feature Selection for Remote Sensing Scene Classification}, 
  volume={12},
  number={11},
  pages={2321-2325},
  year={2015}
}

@article{kingma2014adam,
  title={Adam: A method for stochastic optimization},
  author={Kingma, Diederik P},
  journal={arXiv preprint arXiv:1412.6980},
  year={2014}
}

@inproceedings{gao2023implicit,
  title={Implicit diffusion models for continuous super-resolution},
  author={Gao, Sicheng and Liu, Xuhui and Zeng, Bohan and Xu, Sheng and Li, Yanjing and Luo, Xiaoyan and Liu, Jianzhuang and Zhen, Xiantong and Zhang, Baochang},
  booktitle={Proceedings of the IEEE/CVF conference on computer vision and pattern recognition},
  pages={10021--10030},
  year={2023}
}

@article{zhang2024,
  author={Zhang, Yan and Liu, Hanqi and Li, Zhenghao and Gao, Xinbo and Shi, Guangyao and Jiang, Jianan},
  journal={IEEE Transactions on Geoscience and Remote Sensing}, 
  title={TCDM: Effective Large-Factor Image Super-Resolution via Texture Consistency Diffusion}, 
  volume={62},
  pages={1-13},
  year={2024},
}

@inproceedings{chen2021learning,
  title={Learning continuous image representation with local implicit image function},
  author={Chen, Yinbo and Liu, Sifei and Wang, Xiaolong},
  booktitle={Proceedings of the IEEE/CVF conference on computer vision and pattern recognition},
  pages={8628--8638},
  year={2021}
}

@inproceedings{wang2018esrgan,
  title={Esrgan: Enhanced super-resolution generative adversarial networks},
  author={Wang, Xintao and Yu, Ke and Wu, Shixiang and Gu, Jinjin and Liu, Yihao and Dong, Chao and Qiao, Yu and Change Loy, Chen},
  booktitle={Proceedings of the European conference on computer vision (ECCV) workshops},
  pages={0--0},
  year={2018}
}

@article{brock2018large,
  title={Large Scale GAN Training for High Fidelity Natural Image Synthesis},
  author={Brock, Andrew},
  journal={arXiv preprint arXiv:1809.11096},
  year={2018}
}

@article{hsu2024drct,
  title={DRCT: Saving Image Super-resolution away from Information Bottleneck},
  author={Hsu, Chih-Chung and Lee, Chia-Ming and Chou, Yi-Shiuan},
  journal={arXiv preprint arXiv:2404.00722},
  year={2024}
}

@inproceedings{zhou2023learning,
  title={Learning correction filter via degradation-adaptive regression for blind single image super-resolution},
  author={Zhou, Hongyang and Zhu, Xiaobin and Zhu, Jianqing and Han, Zheng and Zhang, Shi-Xue and Qin, Jingyan and Yin, Xu-Cheng},
  booktitle={Proceedings of the IEEE/CVF International Conference on Computer Vision},
  pages={12365--12375},
  year={2023}
}

@inproceedings{dai2019second,
  title={Second-order attention network for single image super-resolution},
  author={Dai, Tao and Cai, Jianrui and Zhang, Yongbing and Xia, Shu-Tao and Zhang, Lei},
  booktitle={Proceedings of the IEEE/CVF conference on computer vision and pattern recognition},
  pages={11065--11074},
  year={2019}
}

@article{zhu2017deep,
  title={Deep learning in remote sensing: A comprehensive review and list of resources},
  author={Zhu, Xiao Xiang and Tuia, Devis and Mou, Lichao and Xia, Gui-Song and Zhang, Liangpei and Xu, Feng and Fraundorfer, Friedrich},
  journal={IEEE geoscience and remote sensing magazine},
  volume={5},
  number={4},
  pages={8--36},
  year={2017}
}

@article{bashir2021comprehensive,
  title={A comprehensive review of deep learning-based single image super-resolution},
  author={Bashir, Syed Muhammad Arsalan and Wang, Yi and Khan, Mahrukh and Niu, Yilong},
  journal={PeerJ Computer Science},
  volume={7},
  pages={e621},
  year={2021}
}

@article{wang2022comprehensive,
  title={A comprehensive review on deep learning based remote sensing image super-resolution methods},
  author={Wang, Peijuan and Bayram, Bulent and Sertel, Elif},
  journal={Earth-Science Reviews},
  volume={232},
  pages={104110},
  year={2022},
}

@article{dong2021rrsgan,
  title={RRSGAN: Reference-based super-resolution for remote sensing image},
  author={Dong, Runmin and Zhang, Lixian and Fu, Haohuan},
  journal={IEEE Transactions on Geoscience and Remote Sensing},
  volume={60},
  pages={1--17},
  year={2021},
  publisher={IEEE}
}

@inproceedings{dong2024building,
  title={Building bridges across spatial and temporal resolutions: Reference-based super-resolution via change priors and conditional diffusion model},
  author={Dong, Runmin and Yuan, Shuai and Luo, Bin and Chen, Mengxuan and Zhang, Jinxiao and Zhang, Lixian and Li, Weijia and Zheng, Juepeng and Fu, Haohuan},
  booktitle={Proceedings of the IEEE/CVF Conference on Computer Vision and Pattern Recognition},
  pages={27684--27694},
  year={2024}
}

@article{wang2025semantic,
  title={Semantic guided large scale factor remote sensing image super-resolution with generative diffusion prior},
  author={Wang, Ce and Sun, Wanjie},
  journal={ISPRS Journal of Photogrammetry and Remote Sensing},
  volume={220},
  pages={125--138},
  year={2025},
  publisher={Elsevier}
}

@inproceedings{wu2023hsr,
  title={HSR-Diff: Hyperspectral image super-resolution via conditional diffusion models},
  author={Wu, Chanyue and Wang, Dong and Bai, Yunpeng and Mao, Hanyu and Li, Ying and Shen, Qiang},
  booktitle={Proceedings of the IEEE/CVF International Conference on Computer Vision},
  pages={7083--7093},
  year={2023}
}

@article{wang2004image,
  title={Image quality assessment: from error visibility to structural similarity},
  author={Wang, Zhou and Bovik, Alan C and Sheikh, Hamid R and Simoncelli, Eero P},
  journal={IEEE transactions on image processing},
  volume={13},
  number={4},
  pages={600--612},
  year={2004},
  publisher={IEEE}
}

@inproceedings{zhang2018unreasonable,
  title={The unreasonable effectiveness of deep features as a perceptual metric},
  author={Zhang, Richard and Isola, Phillip and Efros, Alexei A and Shechtman, Eli and Wang, Oliver},
  booktitle={Proceedings of the IEEE conference on computer vision and pattern recognition},
  pages={586--595},
  year={2018}
}
}

\end{document}